\documentclass[12pt]{article}
\usepackage{epsfig}
\usepackage{a4wide}

\usepackage{amssymb}
\newcommand{\be}{\begin{equation}}
\newcommand{\ee}{\end{equation}}

\newcommand{\bea}{\begin{eqnarray}}
\newcommand{\eea}{\end{eqnarray}}
\newcommand{\GeV}{\hbox{\rm\,GeV}}
\newcommand{\TeV}{\hbox{\rm\,TeV}}

\def\tm{\vec\tau_-}
\def\tp{\vec\tau_+}

\begin{document}

\begin{flushright}
MZ-TH/09-29\\
arXiv:0908.2027 [hep-ph]\\
August 2009
\end{flushright}

\begin{center}
{\Large\bf Probing scalar particle and unparticle couplings\\[10pt]
  in ${\bf e^+e^-\to t\bar t}$ with transversely polarized beams}\\[24pt]
{\large S.~Groote$^{1,2}$, H.~Liivat$^1$, I.~Ots$^1$ and T.~Sepp$^1$}\\[12pt]
$^1$ Loodus- ja Tehnoloogiateaduskond, F\"u\"usika Instituut,\\
  Tartu \"Ulikool, Riia~142, 51014 Tartu, Estonia\\[12pt]
$^2$ Institut f\"ur Physik der Johannes-Gutenberg-Universit\"at,\\
  Staudinger Weg 7, 55099 Mainz, Germany
\vspace{12pt}
\end{center}


\begin{abstract}
In searching for indications of new physics scalar particle and unparticle
couplings in $e^+e^-\to t\bar t$, we consider the role of transversely
polarized initial beams at $e^+e^-$ colliders. By using a general relativistic
spin density matrix formalism for describing the particles spin states, we
find analytical expressions for the differential cross section of the process
with $t$ or $\bar t$ polarization measured, including the anomalous coupling
contributions. Thanks to the transversely polarized initial beams these
contributions are first order anomalous coupling corrections to the Standard
Model (SM) contributions. We present and analyse the main features of the SM
and anomalous coupling contributions. We show how differences between SM and
anomalous coupling contributions provide means to search for anomalous
coupling manifestations at future $e^+e^-$ linear colliders. 
\end{abstract}

\newpage

\section{Introduction}
The top quark is by far the heaviest fundamental particle. Because of this,
couplings including the top quark are expected to be more sensitive to
new-physics manifestations than couplings to other particles. This is why top
quark physics is a very fascinating field of investigation and has been
developed actively for a long time. During the last decade theoretical
investigations have been connected closely to the physics of near-future
colliders like the Large Hadronic Collider (LHC) at CERN and the International
Linear Collider (ILC). As a matter of fact, the LHC is no longer a future
collider. The setup has been completed and first useful scientific information
will be available in the near future. The center-of-mass energy of $14\TeV$ and
the very large statistics allow one to determine top quark properties
accurately. On the other hand, the future of the ILC is presently unknown.
Nevertheless, we use it as an example of a future $e^+e^-$ linear collider and
its possibilities.

The proposed ILC designed for a center-of-mass starting energy of $500\GeV$
and about three orders less statistics as compared to the LHC is still
considered as a perspective tool for complementary investigations of
new-physics manifestations. The reason is that compared to LHC, the ILC has two
distinctive advantages: a very clean experimental environment and the
possibility to use both longitudinally polarized (LP) and transversely
polarized (TP) beams. Especially the use of TP beams gains more and more
attention. By using LP one can enhance the sensitivity for different parts of
the coupling which, at least in principle, can be measured also for
unpolarized beams. This is because LP does not define additional space
directions. However, TP provide new directions which allow one to analyze
interactions beyond the Standard Model (SM) more efficiently. This facility
should be available at the ILC or other colliders of the same type.
                                           
One of the areas where the advantage of TP beams can be taken is the
investigation of anomalous scalar- and tensor-type couplings. More than thirty
years ago Dass and Ross~\cite{Dass:1975mj} and later
Hikasa~\cite{Hikasa:1985qi} showed that for TP $e^+e^-$ beams the amplitudes
of such couplings interfere with the SM ones. Due to the helicity conservation
this is not the case when using unpolarized or LP beams. For vanishing
initial state masses the scalar- and tensor-type couplings at the $e^+e^-$
vertex are helicity violating, whereas the SM containing vector and
axial vector couplings are helicity conserving. Therefore, in the limit of
massless initial particles there is no nonzero interference terms for
unpolarized and LP beams. However, as the argument of helicity conservation
fails for TP beams, for TP initial beams the scalar--tensor coupling amplitudes
interfere with the SM ones. Ananthanarayan and
Rindani~\cite{Ananthanarayan:2003wi} demonstrated how TP beams can provide
additional means to search for CP violation via interference between SM and
anomalous, scalar--tensor-type coupling contributions in $e^+e^-\to t\bar t$.
Therefore, the use of TP beams enables one to probe new physics appearing
already in first order contributions. In addition, the additional polarization
vector allows one to analyze CP violation asymmetries without the necessity to
final state top or antitop polarizations.

The aforementioned advantages can be used also in analyzing (pseudo)scalar
unparticle manifestations via their virtual effects. The unparticle is a new
concept proposed by Georgi~\cite{Georgi:2007ek} based on the possible
existence of a nontrivial scale-invariant sector with an energy scale much
higher than that of the SM. At lower energies this sector is assumed to couple
to the SM fields via nonrenormalizable effective interactions involving
massless objects of fractional scale dimension $d_u$ coined as unparticles.
Using concepts of effective theories one can calculate the possible effects of
such a scale-invariant sector for $\TeV$-scale colliders. The existence of 
unparticles could lead to measurable deviations from SM predictions as well as
from the predictions of various models beyond SM. The experimental signals of
unparticles might be of two kinds. If unparticles are produced, they manifest
themselves as missing energy and momentum. On the other hand, unparticles can
cause virtual effects in processes of SM particles. 

Since Georgi's significant publications the study of unparticle physics has
gained a lot of attention, shedding light on both theoretical and
phenomenological aspects. The most interesting theoretical developments of
unparticle physics are listed in the introduction of Ref.~\cite{Georgi:2009xq}.
One certainly has to add the content of Ref.~\cite{Georgi:2009xq} written by
Georgi himself and Kats where the self-interaction of unparticles is developed.
The reason is that without self-interaction unparticle physics is incomplete.
Also Ref.~\cite{Georgi:2008pq} of the same authors is of importance here where
the two dimensional toy model of unparticle physics is discussed. The same
model was used for examplifying the methods used in Ref.~\cite{Georgi:2009xq}.
Interesting are also considerations related to the Higgs~\cite{Sannino:2008nv,%
Nelson:2009jg} and to the possibilities of additional observations of CP
violations provided by the unparticle physics (see for instance
Ref.~\cite{Zwicky:2007yc}).

Since unparticle physics has a very rich phenomenology, the number of papers
in this sector is greater than in the theoretical sector. However, it is
difficult to point out the more outstanding ones. A significant part of the
phenomenological studies in particle physics are related to the top quark,
especially to top quark pair production processes in $e^+e^-$ collisions (see
e.g.\ Ref.~\cite{Huitu:2007im} and references therein). A unique feature of
virtual unparticle exchanges is the complex phase of the unparticle propagator
for timelike momenta. If this feature could be identified, it would be a
conclusive device for the existence of unparticles. One way to capture the
feature is again to use TP initial beams at linear $e^+e^-$ collider processes.

In this paper we study how TP initial beams can be used to disentangle scalar
particle and unparticle contributions from SM contributions in the process
$e^+e^-\to t\bar t$. In Sec.~2 we present analytic expression for the
differential cross section of the process with anomalous scalar particle and
virtual scalar unparticle coupling corrections in the case where the top or
antitop quark polarization is measured. In Sec.~3 we present the main features
of the SM, anomalous particle and unparticle contributions and analyze the
methods to isolate signatures for different contributions. Our results can be
used also for analyzing other $e^+e^-$ annihilation processes into
particle--antiparticle pairs. In Sec.~4 we analyze the CP violation effects
caused by the anomalous couplings. In Sec.~5 we consider the possibilities to
use final top (antitop) polarization for disentangling anomalous contributions
from the SM ones. In the last section we draw our conclusions.
 
\section{The differential cross section of the process}
In this section we present the general analytical expressions for the
differential cross section of the process $e^+e^-\to t\bar t$ with arbitrarily
polarized initial beams in the presence of anomalous scalar particle and
scalar unparticle couplings. In doing so we assume that the amplitudes for the
anomalous couplings are much smaller than the amplitudes of SM couplings.
Because of this, the squared amplitude of the SM process can be supplemented
by the interference of SM and anomalous couplings. The electron mass is taken
to be zero. The calculations have been performed in the center-of-mass system
without specifying the coordinate system and spin polarization axes.
 
\subsection{Description of the spin states}
Since the top quark is very heavy, there is no reason to believe that the
helicity basis will be the best choice to describe the top quark polarization
state. Therefore, we use a general relativistic spin density matrix formalism
to describe the particles' spin state. When the top quark polarization is
measured, one replaces  $u(p_t)\bar u(p_t)$ in the squared amplitude by the
density matrix,
\begin{equation}
u(p_t)\bar u(p_t)\ \longrightarrow\ \frac12(\not{\!p}_t+M_t)
  (1+\gamma_5\!\not{\!s}_t)
\end{equation}  
and sums over the spin states of antitop, i.e.\
\begin{equation}
v(p_{\bar t})\bar v(p_{\bar t})\ \longrightarrow\ (\not{\!p}_{\bar t}-M_t).
\end{equation}
If the antitop polarization is measured, one uses the replacements
\begin{eqnarray}
v(p_{\bar t})\bar v(p_{\bar t})&\longrightarrow&\frac12(\not{\!p}_{\bar t}-M_t)
  (1+\gamma_5\!\not{\!s}_{\bar t}),\nonumber \\
u(p_t)\bar u(p_t)&\longrightarrow&(\not{\!p}_t+M_t).
\end{eqnarray}
Here $\not{\!s}=\gamma_\mu s^\mu$ where $s^\mu$ is the polarization four-vector
\begin{equation}
s^\mu=\Bigg(\frac{\vec p\,\vec\xi}{M_t},
  \vec\xi+\frac{(\vec p\,\vec\xi)\vec p}{M_t(E+M_t)}\Bigg),
\end{equation}
and $\vec\xi$ is the polarization vector in the rest frame of the particle
($0\leq|\vec\xi|\leq 1$). Assuming the electron and the positron beams to be
polarized, one replaces both $u(k_-)\bar u(k_-)$ and $v(k_+)\bar v(k_+)$ by
the density matrices 
\begin{eqnarray}
u(k_-)\bar u(k_-)&\longrightarrow&
  \frac12(\not{\!k}_-+m)(1+\gamma_5\!\not{\!s}_-),\nonumber\\
v(k_+)\bar v(k_+)&\longrightarrow&
  \frac12(\not{\!k}_+-m)(1+\gamma_5\!\not{\!s}_+).
\end{eqnarray}
When calculating the process with both LP and TP nonvanishing components, the
limit $m/E\to 0$ can be conveniently taken by making use the
approximation~\cite{Ots:2000pq}
\begin{equation}
s^\mu_\mp\approx\frac{h_\mp k^\mu_\mp}m+\tau_\mp^\mu
\end{equation}
with setting $m=0$ afterwards. $h_\mp$ is the measure of the LP of the
initial beams, $\tau_\mp^\mu=(0,\vec\tau_\mp)$ is the TP four-vector with
$\vec k_\mp\cdot\vec\tau_\mp=0$, and the two signs correspond to the electron
and the positron beam, respectively.

\subsection{Anomalous coupling amplitudes}
We use the effective anomalous scalar\footnote{For simplicity, we use the term
``scalar'' to refer to the combination of scalar and pseudoscalar couplings
used in what follows.} coupling amplitude (particle case) in the form
\begin{equation}\label{Ma}
{\cal M}_p=K_p\,\bar v(k_+)(g_S+ig_P\gamma_5)u(k_-)\,\,
  \bar u(p_t)(c_S+ic_P\gamma_5)v(p_{\bar t}),
\end{equation}
where $g_S$, $g_P$ and $c_S$, $c_P$ are the scalar and pseudoscalar coupling
constants of the electron and the top quark, respectively, and 
$K_p=g_p^2/\Lambda_p^2$ with $g_p$ as a dimensionless coupling constant and
$\Lambda_p$ is the scale of the anomalous scalar particle coupling. In the CM
system one takes $\vec k_-=\vec{k}$, $\vec k_+=-\vec k$, $\vec p_t=\vec p$ and
$\vec p_{\bar t}=-\vec p$. 

The propagator for the scalar unparticle has the general 
form~\cite{Georgi:2007ek,Cheung:2007ap}
\begin{equation}
\Delta=\frac{A_{d_u}}{2\sin(d_u\pi)}(-P^2)^{d_u-2},
\end{equation}
where $d_u$ is the scale dimension and the factor $A_{d_u}$ is given by 
\begin{equation}
A_{d_u}=\frac{16\pi^{5/2}\Gamma(d_u+1/2)}{(2\pi)^{2d_u}
  \Gamma(d_u-1)\Gamma(2d_u)}.
\end{equation} 
In the process under consideration mediated by the $s$-channel unparticle
exchange, the propagator features a complex phase,
\begin{equation}
(-P^2)^{d_u-2}={|P^2|}^{d_u-2}\,e^{-id_u\pi}.
\end{equation}
The Feynman rules for the interaction of the virtual scalar unparticle with
SM fermionic fields can be found in Ref.~\cite{Cheung:2007ap}. We use the
general case with different coupling constants for scalar and pseudoscalar
interactions as well as for different flavors. In this case the virtual
exchange of a scalar unparticle between two fermionic currents can be
expressed by the four-fermion interaction
\begin{equation}\label{Mu}
{\cal M}_u=\frac{g_u^2\,A_{d_u}\,|P^2|^{d_u-2}\,
  e^{-i\,d_u\pi}}{2\sin(d_u\pi)(\Lambda^2_u)^{d_u-1}}
  \bar v(k_+)(g_S+ig_P\gamma_5)u(k_-)\,\,
  \bar u(p_t)(c_S+ic_P\gamma_5)v(p_{\bar t}).
\end{equation} 
In this expression we use the same symbols $g_S$, $g_P$, $c_S$ and $c_P$ for
the scalar and pseudoscalar coupling constants without assuming that they
take the same values as in Eq.~(\ref{Ma}).

\subsection{The expressions for the differential cross section}
Here we present the analytical expressions for the differential cross section
contributed from the three sources: from the SM couplings and from the
interference of the SM couplings with the anomalous scalar (particle)
coupling and scalar unparticle coupling. Each of these three expressions 
describes two cases -- when the top polarization and when the antitop
polarization is measured. All these contributions will be considered in the
following.

\vspace{12pt}
\noindent\underline{The SM couplings}
\begin{equation}\label{int_SM}
\frac{d\sigma_{SM}}{d\Omega}\Big|_{cm}=\frac{p}{256\pi^2k^3}|{\cal M}_{SM}|^2,
\end{equation}
where
\begin{eqnarray}\label{int_sm}
\lefteqn{|{\cal M}_{SM}|^2\ =\ {\cal M}^2_{\gamma\gamma}
  +{\cal M}^2_{ZZ}+2\,\mathrm{Re}\,{\cal M}_\gamma{\cal M}^*_Z}
  \nonumber\\
  &=&8k^2N_C\Bigg\{A_1\left(E^2+p^2\cos^2\theta\right)
  +A_2M^2+4A_3Ep\cos\theta\nonumber\\&&\strut
  -2M\left[\left(A_4\:E+A_6\:p\cos\theta\right)\hat{k}\cdot\vec{s}
  +(A_5\:\frac{p}{E}\cos\theta+A_6)\:\vec{p}\cdot\vec{s}\right]
  \nonumber\\&&\strut
  +A_7\left(-\tm\cdot\tp\:p^2\sin^2\theta
  +2\:\vec{p}\cdot\tm\:\vec{p}\cdot\tp\right)\nonumber\\&&\strut
  - 2A_8\:M\left[\tm\cdot\vec{s}\:\vec{p}\cdot\tp
  +\tp\cdot\vec{s}\:\vec{p}\cdot\tm+\tm\cdot\tp
  \left(-\vec{p}\cdot\vec{s}+\hat{k}\cdot\vec{s}\:p\cos\theta\right)\right]
  \Bigg\},\qquad\quad
\end{eqnarray}
$N_C=3$ is the number of quark colours and
\begin{eqnarray}\label{a1_a8}
A_1&=&K_\gamma^2(1-h_-h_+)
  +K_Z^2(c_V^2+c_A^2)[(g_V^2+g_A^2)(1-h_-h_+)+2g_Vg_A(h_+-h_-)]
  \nonumber\\&&\strut
  +2K_\gamma K_Zc_V[g_V(1-h_-h_+)+g_A(h_+-h_-)]\nonumber\\
A_2&=&K_\gamma^2(1-h_-h_+)
  +K_Z^2(c_V^2-c_A^2)[(g_V^2+g_A^2)(1-h_-h_+)+2g_Vg_A(h_+-h_-)]
  \nonumber\\&&\strut
  +2K_\gamma K_Zc_V[g_V(1-h_-h_+)+g_A(h_+-h_-)]\nonumber\\
A_3&=&K_Z^2c_Vc_A[(g_V^2+g_A^2)(h_+-h_-)+2g_Vg_A(1-h_-h_+)]
  \nonumber\\&&\strut
  +K_\gamma K_Zc_A[g_V(h_+-h_-)+g_A(1-h_-h_+)]\nonumber\\
A_4&=&K_\gamma^2(h_+-h_-)
  +K_Z^2c_V^2[(g_V^2+g_A^2)(h_+-h_-)+2g_Vg_A(1-h_-h_+)]
  \nonumber\\&&\strut
  +2K_\gamma K_Zc_V[g_V(h_+-h_-)+g_A(1-h_-h_+)]\nonumber\\
A_5&=&K_Z^2c_A^2[(g_V^2+g_A^2)(h_+-h_-)+2g_Vg_A(1-h_-h_+)]\nonumber\\
A_6&=&K_Z^2c_Vc_A[(g_V^2+g_A^2)(1-h_-h_+)+2g_Vg_A(h_+-h_-)]
  \nonumber\\&&\strut
  +K_\gamma K_Zc_A[g_V(1-h_-h_+)+g_A(h_+-h_-)]\nonumber\\
A_7&=&K_\gamma^2
  +K_Z^2(c_V^2+c_A^2)(g_V^2-g_A^2)+2K_\gamma K_Zc_Vg_V\nonumber\\
A_8&=&K_Z^2c_Vc_A(g_V^2-g_A^2)+K_\gamma K_Zc_Ag_V
\end{eqnarray}
with 
\begin{equation}\label{KgammaKZ}
K_\gamma=\frac{Q_fe^2}{4k^2},\qquad
K_Z=-\frac{e^2}{\sin^2(2\theta_W)(4k^2-M^2_Z)}.
\end{equation}
We use the three LP-dependent coefficients $A_i$ ($i=1,2,3$) for the
unpolarized final state and the three LP-dependent coefficients $A_i$
($i=4,5,6$) for the polarized final state. The two coefficients $A_i$
($i=7,8$) which do not depend on the LP parameters $h_\pm$ are used for
contributions which depend on the initial state transverse polarization for
unpolarized ($A_7$) and polarized final state ($A_8$). The coefficients are
used to disentangle the coupling constants and LP parameters from the
kinematical parts as much as possible. $g_V$, $g_A$ and $c_V$, $c_A$ are the
vector and axial vector coupling constants of the electron and the top quark,
respectively, and $Q_f=+2/3$ is the electric charge of the top quark.
$\hat{k}=\vec{k}/k$ is the unit vector given by the momentum $\vec{k}$, and
$k=|\vec{k}|$ is the energy of the electron. $E=k$ is the top quark energy,
and $\vec{p}\/$ is the momentum of the top quark ($p=\sqrt{E^2-M^2}$).
Finally, $\theta$ and $\vec{s}$ are the scattering angle (with
$\cos\theta=\hat{k}\cdot\vec{p}/p$) and the polarization vector of the top
quark. Both the top and antitop polarization measured cases have been
described in terms of the top momentum and scattering angle. We have also used
the same notation $\vec{s}$ for the top and antitop polarization vectors
$\vec{s}_t$ and $\vec{s}_{\bar{t}}$.  The polarization quantities $h_\pm$ and
$\tau_\pm$ of the initial beam are defined in Sec.~2.1. As a result the
expressions in top and antitop polarization measured cases entirely coincide.
If one would like to describe the antitop case in terms of antitop parameters,
one has to take $\vec{p}$ and $\cos\theta$ with the opposite signs
($-\vec{p},\: -\cos\theta$). This procedure changes the signs in front of a
part of the terms in Eq.~(\ref{int_sm}) and one has to use upper and lower
signs to distinguish top and antitop cases.

We have not used the Mandelstam variables because this makes the expressions
cumbersome and less clear for their further analysis. For the same reason we
have not expressed the top quark's energy and momentum by the energy and
momentum of the electron.

\vspace{12pt}
\noindent\underline{The interference of the SM and
  anomalous scalar particle coupling}\\
In taking into account anomalous scalar and pseudoscalar contributions, we
can assume that the factor $K_p$ is small. Therefore, we can skip the
contribution $|{\cal M}_p|^2$ and obtain
\begin{equation}\label{approx_scalar}
\frac{d\sigma_{SM+p}}{d\Omega}\Big|_{cm}\
  \approx\ \frac{d\sigma_{SM}}{d\Omega}\Big|_{cm}
  +\frac{p}{256\pi^2k^3}K_p{\cal B}_a,
\end{equation}
where the momenta $p$ and $k$ are defined in the paragraph following
Eq.~(\ref{KgammaKZ}), and
\begin{eqnarray}\label{int_scalar}
 \lefteqn{{\cal B}_a\ =\ 2\mathrm{Re}{\cal M}_{SM}{\cal M}_p^*/K_p
  \ =}\nonumber\\
  &=&-16k^2N_C\bigg\{\left(K_\gamma+K_Zg_Vc_V\right)
  \Big[g_S\:\hat{k}\times(\tm+\tp)-g_P\:(\tm-\tp)\Big]
  \cdot\strut\nonumber\\&&\strut\cdot
  \Big[c_SE\vec{p}\times\vec{s}
  \pm c_P(\vec{p}\cdot\vec{s}\:\vec{p}-E^2\vec{s})\Big]
  +K_Zg_A\Big[g_S(\tm+\tp)+g_P\hat{k}\times(\tm-\tp)\Big]
  \cdot\strut\nonumber\\&&\strut\cdot
  \Big[c_Vc_S\:M\vec{p}\pm c_Ac_P\:E\vec{p}\times\vec{s}
  -c_Ac_S(\vec{p}\cdot\vec{s}\:\vec{p}-p^2\vec{s})\Big]
  +\strut\nonumber\\&&\strut
  +\Big[g_S(h_+\tm-h_-\tp)+g_P\hat{k}\times(h_+\tm+h_-\tp)\Big]
  \cdot\strut\nonumber\\&&\strut\cdot
  \Big[(K_\gamma+K_Zg_Vc_V)c_SM\vec{p}+K_Zg_Vc_A(\pm c_PE\vec{p}\times\vec{s}
  -c_S(\vec{p}\cdot\vec{s}\,\vec{p}-p^2\vec{s}))\Big]
  +\strut\nonumber\\&&\strut
  +K_Zg_Ac_V\Big[g_S\hat{k}\times(h_+\tm-h_-\tp)-g_P(h_+\tm+h_-\tp)\Big]
  \cdot\strut\nonumber\\&&\strut\cdot
  \Big[c_SE\vec{p}\times\vec{s}
  \pm c_P(\vec{p}\cdot\vec{s}\,\vec{p}-E^2\vec{s})\Big]\bigg\}.
\end{eqnarray}

\vspace{12pt}
\noindent\underline{The interference of the SM and
  anomalous scalar unparticle coupling}\\
Apart from the different overall constants, the real part of the complex phase
in the unparticle amplitude in Eq.~(\ref{Mu}) leads to the same expression
${\cal B}_a$ as the scalar particle coupling amplitude. Therefore, one can
write
\begin{equation}\label{approx_unparticle}
\frac{d\sigma_{SM+u}}{d\Omega}\Big|_{cm}
  \ \approx\ \frac{d\sigma_{SM}}{d\Omega}\Big|_{cm}
  +\frac{p}{256\pi^2k^3}K_u
  \left(\cos(d_u\pi){\cal B}_a+\sin(d_u\pi){\cal B}_b\right),
\end{equation}
where ${\cal B}_a$ is given in Eq.~(\ref{int_scalar}) and
\begin{eqnarray}\label{int_unparticle}
\lefteqn{{\cal B}_b\ =\ 
  2{\mathrm Re}{\cal M}_{SM}{\cal M}_u^*/K_u
  \ =\ -16k^2N_C\bigg\{\left[g_S\hat{k}\times(\tm+\tp)-g_P(\tm-\tp)\right]
  \cdot\strut}\nonumber\\&&\strut\cdot
  \Big[(K_\gamma+K_Zc_Vg_V)c_SM\vec{p}
  +K_Zc_Ag_V\left(\pm c_P E\vec{p}\times\vec{s}
  -c_S(\vec{p}\cdot\vec{s}\,\vec{p}-p^2\vec{s})\right)\Big]
  +\strut\nonumber\\&&\strut
  -K_Zc_Vg_A\Big[g_S(\tm+\tp)+g_P\hat{k}\times(\tm-\tp)\Big]\cdot
  \Big[c_SE\vec{p}\times\vec{s}
  \pm c_P(\vec{p}\cdot\vec{s}\:\vec{p}-E^2\vec{s})\Big]
  +\strut\nonumber\\&&\strut
  -(K_\gamma+K_Zg_Vc_V)
  \Big[g_S(h_+\tm-h_-\tp)+g_P\hat{k}\times(h_+\tm+h_-\tp)\Big]
  \cdot\nonumber\\&&\strut\cdot
  \Big[c_SE\vec{p}\times\vec{s}
  \pm c_P(\vec{p}\cdot\vec{s}\:\vec{p}-E^2\vec{s})\Big]
  +\strut\nonumber\\&&\strut
  +K_Zg_A\Big[g_S\hat{k}\times(h_+\tm-h_-\tp)-g_P(h_+\tm+h_-\tp)\Big]
  \cdot\strut\nonumber\\&&\strut\cdot
  \Big[c_Vc_SM\vec{p}\pm c_Ac_PE\vec{p}\times\vec{s}
  -c_Ac_S(\vec{p}\cdot\vec{s}\:\vec{p}-p^2\vec{s})\Big]\bigg\},
\end{eqnarray}
where
\begin{equation}
K_u=\frac{g_u^2\,A_{d_u}\,|P^2|^{d_u-2}}{2\sin(d_u\pi)(\Lambda^2_u)^{d_u-1}}
\end{equation}
(note that $|P^2|=4k^2$). In the following we consider the region $1\le d_u<2$.
For $d_u=1$ the unparticle contribution is given by ${\cal B}_a$ which is
already used in the contribution of the anomalous scalar and pseudoscalar
particle interactions. On the other hand, for $d_u=3/2$ the contribution is
given purely by ${\cal B}_b$. However, we do not restrict to these two values
but consider the whole interval.

\section{The main features of the contributions}
Using the approximation in Eq.~(\ref{approx_scalar})
or~(\ref{approx_unparticle}) where the squared amplitude $|{\cal M}_p|^2$
resp.\ $|{\cal M}_u|^2$ is neglected, the process $e^+e^-\to t\bar t$ is fully
described by the analytical expressions for the SM and the anomalous scalar
particle and unparticle coupling contributions. In this section we report
about observations on the SM and anomalous coupling contributions. We present
and analyze the main features of the contributions and the differences between
the SM and anomalous coupling contributions as well as between the scalar
particle and unparticle coupling contributions. These differences are helpful
in disentangling the different contributions at future $e^+e^-$ colliders.
Part of the features given below are already known. We present them for
completeness only.

\subsection{Standard Model versus anomalous coupling contributions}
In comparing the SM contribution with the contribution from the scalar
particle or unparticle coupling, we come to the following conclusions:

\begin{enumerate}

\item The SM contributions depend on the longitudinal polarization of the
initial beams through the coefficients $A_i$ ($i=1,\ldots,6$) which contains
the LP parameters $h_-$ and $h_+$ as well as the coupling constants ($g_V$,
$g_A$, $c_V$ and $c_A$). The coefficients $A_i$ contain both linear and
quadratic terms in the LP parameters. By changing the values of $h_-$ and
$h_+$ one can substantially increase or decrease the coefficients $A_i$ and
by this selected parts of the coupling. However, one cannot form observables 
different from those of the unpolarized beams. The anomalous scalar 
(particle and unparticle) coupling contributions depend linearly on the
longitudinal polarization. However, the LP-dependent terms cannot occur without
the existence of TP vectors: the LP parameters $h_-$ and $h_+$ are always
multiplied by the vectors $\tm$, $\tp$ in combinations $h_-\tp$ and $h_+\tm$. 

\item In the SM contributions the TP-dependent terms depend quadratically
on the TP vectors. Due to this they are different from zero only when both of
the initial beams have TP components. In the anomalous coupling contributions
all the terms have to be and are TP-dependent. They depend linearly on the TP
vectors without or with the multiplicative LP parameters and, as a consequence,
can be different from zero also in the case where only one of the initial beams
is transversely polarized. The linear dependence provides a crucial tool at
future $e^+e^-$ linear colliders for isolating signatures of anomalous scalar
couplings from the SM ones. 

\item In the SM contributions all the terms depending on the final state
polarizations are proportional to the final state fermion mass while the terms
independent of the final state polarizations for the most part are independent
of this mass. On the other hand, for the anomalous coupling contributions the
term independent of the final state polarization is proportional to the final
state fermion mass which is not the case for most of the terms depending on
the final state polarization. This fact stresses the advantages of
investigating final state polarization effects in $e^+e^-$ annihilation just
for top-antitop pair productions. 

\item The SM contributions are invariant with respect to the interchange
$\tm\rightleftarrows\tp$. Since applying the CP transformation to the process
causes the same changes, the above given invariance once more reflects CP
conservation at that level. On the other hand, both the scalar particle and
unparticle contributions contain CP-odd (weak phase) contributions due to the
nonvanishing pseudoscalar coupling constants $g_P$ and $c_P$ in
Eqs.~(\ref{Ma}) and~(\ref{Mu}), respectively. The contributions depend on
the TP vectors through the four combinations
\begin{eqnarray}
&\left[g_S\hat{k}\times(\tm+\tp)-g_P(\tm-\tp)\right],&\label{cnt1}\\
&\left[g_S(\tm+\tp)+g_P\hat{k}\times(\tm-\tp)\right],&\label{cnt2}\\
&\left[g_S(h_+\tm-h_-\tp)+g_P \hat{k}\times(h_+\tm+h_-\tp)\right],
  &\label{cnt3}\\
&\left[g_S\hat{k}\times(h_+\tm-h_-\tp)-g_P(h_+\tm +h_-\tp)\right].
  &\label{cnt4}  
\end{eqnarray}
The fact that under CP the TP vectors $\tm$ and $\tp$ interchange suggests
that there have to be CP-odd terms in the anomalous coupling contributions and
that CP invariance is violated in the process. 

\item Expressing the results in terms of the momentum and scattering angle of
the top quark, the SM contributions to the differential cross section is
independent on whether the top or antitop polarization $\vec s$ is measured.
This is not the case for the anomalous contributions. Here the terms
containing the coupling constant $c_P$ have opposite signs for the case of top
and antitop polarization measurement. As we will see later, this leads to
different CP-odd parts in the $c_P$- and $c_S$-dependent terms.  

\item The TP-dependent terms of the SM contributions vanish at the threshold
of the process. On the other hand, in the anomalous coupling contributions
there exist terms that survive at the threshold. This gives an additional tool
for separating anomalous coupling contributions from the SM ones.

\end{enumerate}

\subsection{Scalar particle versus unparticle coupling contributions}
In comparing the contributions including scalar particle and unparticle
couplings, we obtain the following conclusions: 

\begin{enumerate}

\item The scalar particle and unparticle coupling contributions in
Eqs.~(\ref{int_scalar}) and~(\ref{int_unparticle}) depend on the same
combinations of TP vectors $\tm$ and $\tp$ as given in
Eqs.~(\ref{cnt1})--(\ref{cnt4}). This makes it difficult to separate these
contributions.
 
\item However, in the scalar particle and the unparticle coupling
contributions the TP-dependent combinations in Eqs.~(\ref{cnt1})--(\ref{cnt4})
are multiplied by different final state expressions. In principle, this
enables us to disentangle the different contributions by measuring the final
state polarizations.

\end{enumerate}

\section{CP violation analysis}
CP violation in weak interactions was first reported for the neutral $K$-meson
system~\cite{Christenson:1964fg}. Further examples were found for $D$- and
$B$-meson systems~\cite{Miyake:2005qb,Aubert:2005rn}. Apart from this, the CP
violation due to SM interactions is predicted to be unobservably
small~\cite{Ananthanarayan:2003wi,Blinov:2008mu}. Hence, one of the
important indications of new physics would be the observation of CP violation
outside the aforementioned systems.

In this section we demonstrate that due to anomalous scalar particle or
unparticle coupling corrections to the SM contribution the CP symmetry in
$e^+e^-\to t\bar{t}$ is violated. We investigate how the interference between
SM and anomalous couplings gives rise to CP-odd quantities in case of
transversely polarized initial beams and construct the CP-odd asymmetries
sensitive to CP violation. For testing CP violation in the process it is not
sufficient to measure only the momenta $\vec{k}$ and $\vec{p}$ because the
only scalar observable which can be constructed from these vectors is
$\vec{k}\cdot\vec{p}$ which is CP-even. Therefore, either initial or final
state polarization vectors are needed. In the case under consideration the TP
initial beams are mandatory: the interference between SM and scalar anomalous
couplings are nonvanishing only with TP initial beams. The possibility to
test CP violation in $e^+e^-\to t\bar{t}$ with TP beams in the presence of
scalar- and tensor-type anomalous couplings was first demonstrated in
Ref.~\cite{Ananthanarayan:2003wi} without measured final state polarization.

The SM and anomalous contributions given by Eqs.~(\ref{int_sm}),
(\ref{int_scalar}) and~(\ref{int_unparticle}) enable to construct CP-odd
asymmetries for transversely polarized initial beams both in the case of
observed and nonobserved final top (antitop) polarization for scalar particle
and unparticle interactions. For both initial beams transversely polarized,
we take $h_-=h_+=0$.

\subsection{CP violation for unpolarized final state quarks}
\begin{figure}[ht]
\vspace{1.5truecm}\centering
\epsfig{file=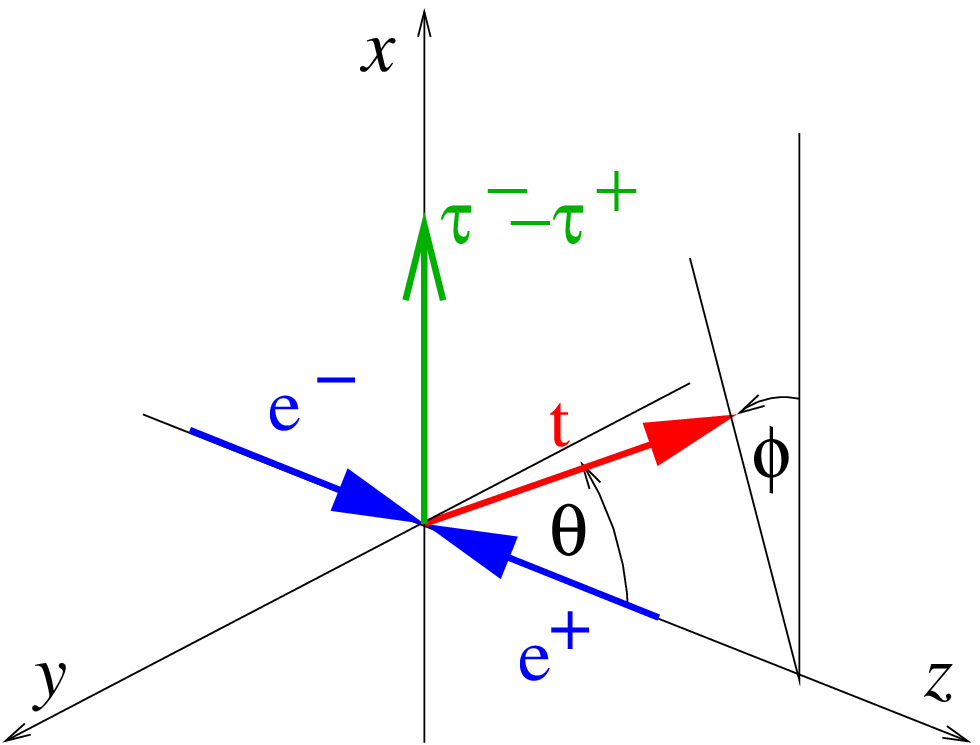, scale=0.7}
\hspace{0.5truecm}
\epsfig{file=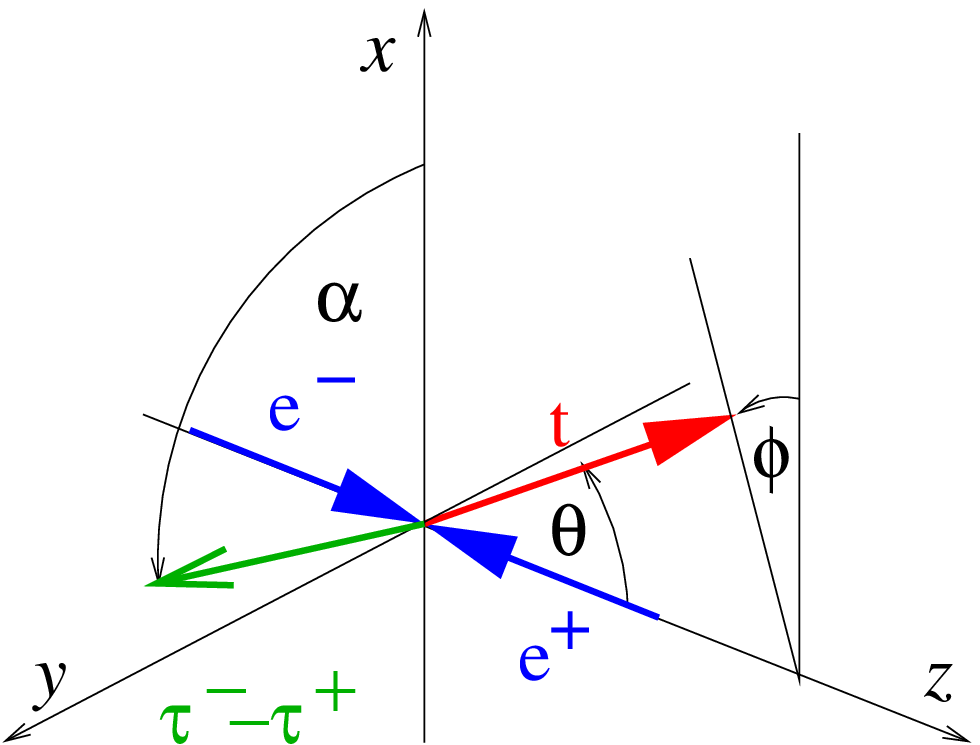, scale=0.7}
\vspace{0.5truecm}
\centerline{\large(a)\kern204pt(b)}
\caption{\label{figure1}Choice of the kinematics for the analysis of scalar
particle (a) and unparticle interactions (b) with
$\tan\alpha=-({\cal A}_a/{\cal A}_b)\tan(d_u\pi)$, where ${\cal A}_{a,b}$
are given in Eq.~(\ref{tanalpha}).}
\end{figure}

Let us first consider the case where the final particle spin states are not
observed. In this case both the scalar particle and unparticle coupling
contributions in Eqs.~(\ref{int_scalar}) and~(\ref{int_unparticle}) do not
depend on the final state top (antitop) polarization vector $\vec s$. In
taking $h_-=h_+=0$ as proposed, only a single term remains in both
contributions ${\cal B}_a$ and ${\cal B}_b$. For the particle coupling
contribution this term contains the TP-dependent factor~(\ref{cnt2}),
\begin{equation}\label{CP_particle}
{\cal B}_a=-16k^2N_CK_Zc_Vg_A[g_S(\tm+\tp)+g_P\hat k\times(\tm-\tp)]\cdot
  c_SM\vec p.
\end{equation}
As applied to the process, the CP transformation interchanges the TP vectors
of the electron and the positron whereas the momenta $\vec k$ and $\vec p$
remain unchanged. As a consequence, the second part of the contribution
in Eq.~(\ref{CP_particle}) depending on the difference $(\tm-\tp)$ changes
sign under the CP transformation (i.e.\ it is CP-odd). Therefore, CP is
violated in the process. One can construct an asymmetry which is sensitive to
CP violation in case where the TP vectors $\tm$ and $\tp$ of the electron and
positron have opposite directions. If we use a coordinate system where the
$z$-axis is determined by the electron momentum $\vec k$, we can direct the
$x$-axis along the electron and opposed to the positron TP vectors. The
situation is illustrated in Fig.~\ref{figure1}(a). Such a choice leads to the
CP-odd quantity
\begin{equation}\label{sin_sin_1}
\frac{\hat{k}\times(\tm-\tp)\cdot\vec{p}}{|\tm-\tp|p}=\sin\theta\sin\phi
\end{equation}
in the differential cross section, where $\phi$ is the azimuthal angle of the
process.

\vspace{7pt}
In the unparticle case the contribution
$\cos(d_u\pi){\cal B}_a+\sin(d_u\pi){\cal B}_b$ with
\begin{equation}\label{CP_unparticle}
{\cal B}_b=-16k^2N_C(K_\gamma+K_Zc_Vg_V)
  [g_S\hat k\times(\tm+\tp)-g_P(\tm-\tp)]\cdot c_SM\vec p
\end{equation}
contains both TP-dependent factors~(\ref{cnt1}) and~(\ref{cnt2}), mixed by the
angle $d_u\pi$. The part in (\ref{cnt1}) causing CP violation in the process
is $g_P(\tm-\tp)\cdot\vec p$. If the unparticle dimension $d_u$ is given by a
specific model, the CP-odd quantity in the differential cross section
corresponding to Eq.~(\ref{sin_sin_1}) is achieved when vectors $\tm$ and
$\tp$ are taken to be opposite and directed along an axis which is rotated by
$\alpha$ with
\begin{equation}\label{tanalpha}
\tan\alpha=-\frac{{\cal A}_a}{{\cal A}_b}\tan(d_u\pi),\qquad
{\cal A}_a:=K_\gamma+K_Zc_Vg_V,\quad
{\cal A}_b:=K_Zc_Vg_A
\end{equation}
starting from the positive and negative direction of the $x$-axis,
respectively (cf.\ Fig.~\ref{figure1}(b)). In this case we obtain a CP-odd
quantity
\begin{equation}\label{sin_sin_2}
\frac{[\hat k\times(\tm-\tp)\cos\alpha+(\tm-\tp)\sin\alpha]\cdot
  \vec{p}}{|\tm-\tp|p}=\sin\theta\sin\phi.
\end{equation}
In both cases one can construct
the CP-odd asymmetry
\begin{equation}\label{asym}
{\cal A}({\theta})=\frac{\displaystyle\int_0^\pi\frac{d\sigma}{d\Omega}d\phi
  -\int_\pi^{2\pi}\frac{d\sigma}{d\Omega}d\phi}{\displaystyle\int_0^\pi
  \frac{d\sigma}{d\Omega}d\phi+\int_\pi^{2\pi}\frac{d\sigma}{d\Omega}d\phi},
\end{equation}
where $\sigma=\sigma_{SM+p}$ or $\sigma_{SM+u}$, resp. Such a quantity for the
scalar- and tensor-type (particle) couplings was first constructed and
analyzed by Ananthanarayan and Rindani~\cite{Ananthanarayan:2003wi}. They
estimated the sensitivity of planned future colliders to new-physics CP
violation in $e^+e^-\to t\bar t$ and showed the possibility to put a bound of
approx.\ $7\TeV$ on the new-physics scale.

\subsection{CP violation and final state polarization}
CP-odd contributions are also observed in the terms which depend on the final
state top or antitop polarizations. These polarizations can be determined by
analyzing the distributions of the final state charged leptons from the top
(or antitop) decay. This method is viable in the top quark case because the
top quark is so massive that it decays before it can hadronize, therefore
avoiding masking nonperturbative effects. Of course, the observation of the CP
violation through the measurement of the final state polarization means a loss
of statistics. On the other hand, this shortage might be partly softened by
the fact that most of the polarization depending terms are not proportional to
the top mass and can be quite large as compared to terms independent of the
top polarization.

If we divide the polarization vector $\vec{s}$ of the final top or antitop
quark into a longitudinal and a transverse part,
\begin{equation} 
\vec{s}=\vec{s}_L+\vec{s}_T=\frac{E}{Mp}h\vec{p}+\vec{\tau},
\label{toppolvec}
\end{equation}
there is only a single term in both ${\cal B}_a$ and ${\cal B}_b$ that depends
on $\vec{s}_L$. While $\vec{p}\cdot\vec{s}_L\vec{p}-p^2\vec{s}_L=0$, the factor
\begin{equation}
\vec{p}\cdot\vec{s}_L\vec{p}-E^2\vec{s}_L=-\frac{EMh}p\vec{p}
\end{equation}
is proportional to the top mass. This factor is multiplied by the $\tp$- and
$\tm$-dependent expressions~(\ref{cnt1}) in the particle and both~(\ref{cnt1})
and~(\ref{cnt2}) in the unparticle case. Besides this, the terms containing
these factors are $c_P$-dependent and therefore, as mentioned in point~5 of
Sec.~3.1, have different signs if both the contributions from top and antitop
polarization measurements are given by the top parameters ($\vec{p},\theta$).
Due to this the CP-odd terms depend on $(\tm+\tp)$ in the combinations
$h\hat{k}\times(\tm+\tp)\cdot\vec{p}$ for the particle case and in addition on
$h(\tm+\tp)\cdot\vec{p}$ for the unparticle case.

The terms depending on the transverse polarization of the final top (antitop)
are not proportional to the top mass. One can divide these terms into
$c_S$-dependent and $c_P$-dependent parts. In the $c_S$-dependent terms the
CP-odd parts depend on the difference of the $\tm$ and $\tp$ vectors while in
the $c_P$-dependent parts they depend on the sum of these vectors. However,
the CP-odd parts in the corresponding terms of scalar particle and scalar
unparticle contributions depend differently on these vectors. If the CP-odd
part of some scalar particle contribution term contains the factor $\tm-\tp$ 
(or $\tm+\tp$), the corresponding term in unparticle case depends in addition
on $\hat{k}\times(\tm-\tp)$ (or $\hat{k}\times(\tm+\tp)$) and {\it vice versa}.
This circumstance might enable one, at least in principle, to separate CP-odd
asymmetries in scalar particle and unparticle cases.

\section{Final state polarizations}
In this section we consider the actual polarizations of the final top or
antitop quarks. The final quark polarizations provide additional tools for
studying the mechanisms of the process and for separating the anomalous
coupling contributions from the SM ones. It is well known that in the Born
approximation the process $e^+e^-\to t\bar t$ with unpolarized or
longitudinally polarized initial beams produces final quarks with polarization
vector lying in the scattering plane~\cite{Fischer:1998gsa}. When using TP
beams the TP vectors $\tm$ and $\tp$ move the final top or antitop
polarization vectors out of scattering plain. Therefore, in the approximation
used the deviation of the final quark polarization vectors from the reaction
plain is only due to the TP initial beams. SM contributes to TP-dependent
terms only if both of the initial beams are transversely polarized. If only
one of the initial beams is transversely polarized, such a deviation would
indicate the presence of anomalous couplings.

\subsection{Top polarization for the SM}
\begin{figure}
\vspace{1.5truecm}\centering
\epsfig{file=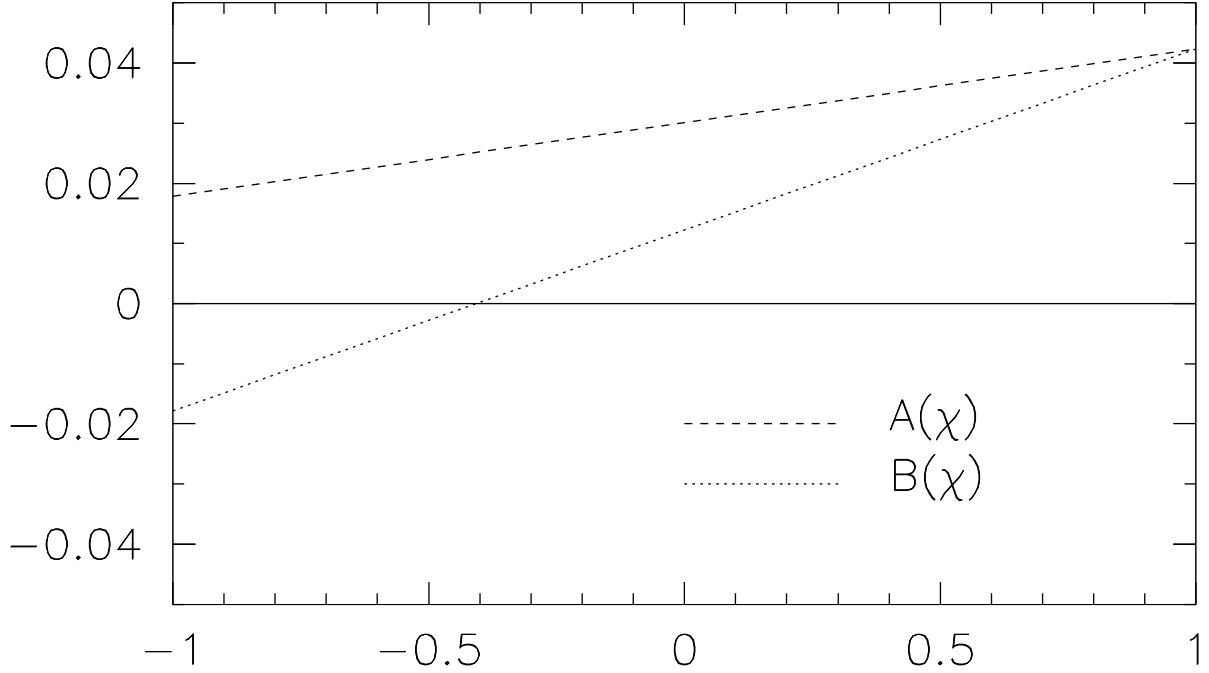, scale=1.0}
\vspace{0.5truecm}
\caption{\label{figure2}The dependence of $A(\chi)$ and $B(\chi)$ on $\chi$.}
\end{figure}

Let us consider the final state polarization in more detail at the threshold of
the process. At threshold the analytical expressions for the differential cross
sections in Eqs.~(\ref{int_sm}), (\ref{int_scalar}) and~(\ref{int_unparticle})
simplify considerably and the polarization properties of the quarks are
displayed more clearly. We start our investigations from the SM sector
considering the polarization properties of the top (antitop) quarks more
generally. Since the TP-dependent terms vanish at the threshold, the main
question will be how much one can tune the top (antitop) quark polarization by
varying the LP parameters $h_+$ and $h_-$ of the initial beams. Indeed, the
result for the polarization turns out to depend effectively on the parameter
\begin{equation}
\chi=\frac{h_+-h_-}{1-h_+h_-}.
\end{equation}
At threshold the squared SM amplitude takes the form
\begin{equation}\label{thressm}
|\mathcal{M}_{SM}|^2|_{\mathrm{thres}}
  =24M^4\left[A_1+A_2-2A_4\hat{k}\cdot\vec{\xi}\right],
\end{equation}
where $\vec{\xi}$ is the top quark polarization vector and $K_\gamma$ 
and $K_Z$ have their threshold forms
\begin{equation}
K_\gamma=\frac{e^2}{6M^2},\qquad
K_Z=-\frac{e^2}{\sin^2(2\theta_W)(4M^2-M_Z^2)}.
\end{equation}
Using the method given in Ref.~\cite{Ots:2000pq} one can find the magnitude
and direction of the actual polarization vector of the top quark,
\begin{equation}\label{smpolvec}
\vec{\xi}_{SM}= -\frac{B(\chi)\hat{k}}{A(\chi)},
\end{equation}
where
\begin{eqnarray}\label{ABdef}
A(\chi)&=&\frac{A_1+A_2}{2(1-h_+h_-)}\ =\ a_1+a_2\chi,\nonumber\\
B(\chi)&=&\frac{A_4}{1-h_+h_-}\ =\ a_1\chi+a_2
\end{eqnarray}
with
\begin{eqnarray}
a_1&=&K_\gamma^2+K_Z^2c_V^2(g_V^2+g_A^2)+2K_\gamma K_Zc_Vg_V,\nonumber\\[7pt]
a_2&=&2K_Zc_Vg_A(K_\gamma+K_Zc_Vg_V).
\end{eqnarray}
In Fig.~\ref{figure2} the dependence of $A(\chi)$ and $B(\chi)$ on $\chi$ is
given.

For the SM sector we use the values of the coupling constants and other
parameters as given by the Particle Data Group~\cite{Amsler:2008zzb},
$g_V=-0.037$, $g_A=-0.5$, $c_V=0.191$, $c_A=0.5$, $g=e/\sin\theta_W$,
$\sin^2\theta_W=0.2415$, $M_t=171.2\GeV$, and $M_Z=91.2\GeV$. We draw the
attention to the fact that for $\chi_0=-0.408$ we obtain $B(\chi_0)=0$.
Therefore, at this value, $\chi=\chi_0$, the top polarization in the process
appears only due to the anomalous coupling contributions. At the same time
$A(\chi_0)$ is smaller than at the point $\chi=0$ and as a consequence the top
polarization from anomalous couplings is larger than in the case of
unpolarized initial beams.

\begin{figure}[t]
\vspace{1.5truecm}\centering
\epsfig{file=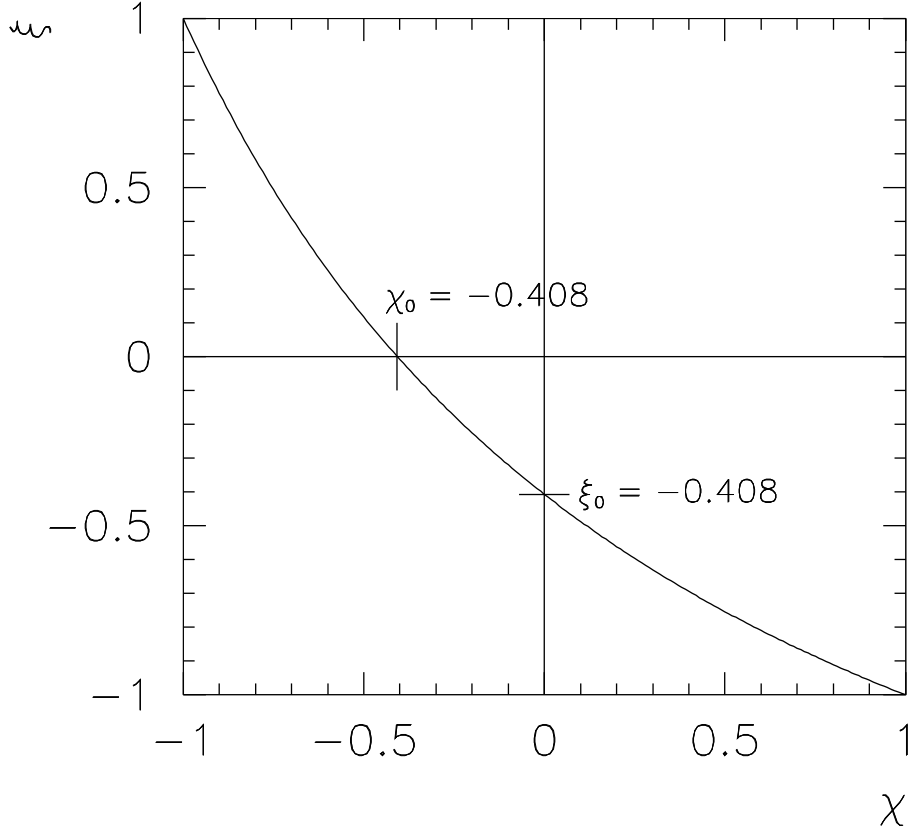, scale=1.0}
\vspace{0.5truecm}
\caption{\label{figure3}The top polarization in SM at threshold
  as a function of $\chi$.}
\end{figure}

Fig.~\ref{figure3} shows how much the top polarization vector can be tuned by
$\chi$ as compared to the case $\chi=0$, where the polarization is given
by~\cite{Harlander:1996vg}
\begin{equation}\label{pol_value}
\vec{\xi}_{SM}|_{\chi=0}=-0.408\hat{k}.
\end{equation}

The fact that the magnitude of the top polarization vector at $\chi=0$ given in
Eq.~(\ref{pol_value}) is equal to the value of $\chi$ at which $\vec{\xi}_{SM}$
vanishes is not an occasional coincidence but a consequence of the special
shape of the structure functions $A(\chi)$ and $B(\chi)$ in
Eqs.~(\ref{ABdef}). The polarization function $\xi_{SM}$ in
Eq.~(\ref{smpolvec}) is of the same shape as the reciprocal function
\begin{equation}
\chi(\xi_{SM})=-\frac{a_1+a_2\xi_{SM}}{a_1\xi_{SM}+a_2}.
\end{equation}
As a consequence, $\xi_{SM}(\chi=0)=\chi(\xi_{SM}=0)=-a_1/a_2=-0.408$.

\subsection{Anomalous coupling corrections to the SM top polarization}
The anomalous scalar particle coupling corrections to the SM contribution at
threshold are given by
\begin{eqnarray}
{\cal B}_a|_{\mathrm{thres}}&=&\pm 48M^4c_P\vec{\xi}\Big\{(K_\gamma+K_Zc_Vg_V)
  [g_S\hat{k}\times(\tm+\tp)-g_P(\tm-\tp)]+\strut\nonumber\\&&\strut\qquad
  +K_Zc_Vg_A[g_S\hat k\times(h_+\tm-h_-\tp)-g_P(h_+\tm+h_-\tp)]\Big\}.
\end{eqnarray}
For the anomalous scalar unparticle corrections one obtains in addition
\begin{eqnarray}
{\cal B}_b|_{\mathrm{thres}}&=&\pm 48M^4c_P\vec{\xi}\Big\{(K_\gamma+K_Zc_Vg_V)
  [g_S(h_+\tm-h_-\tp)+g_P\hat k\times(h_+\tm+h_-\tp)]
  +\strut\nonumber\\&&\strut\qquad
  +K_Zc_Vg_A[g_S(\tm+\tp)+g_P\hat{k}\times(\tm-\tp)]\Big\}.
\end{eqnarray}
For $h_-=h_+=0$, the corresponding corrections to the top polarization vector
are
\begin{equation}
\vec{\xi}_p=K_p\vec{\xi}_a,\qquad
\vec{\xi}_u=K_u(\cos(d_u\pi)\vec{\xi}_a+\sin(d_u\pi)\vec{\xi}_b)
\end{equation}
with
\begin{eqnarray}
\vec{\xi}_a&=&\frac{{\cal A}_ac_P}{A(0)}
  [g_S\hat{k}\times(\tm+\tp)-g_P(\tm-\tp)],\nonumber\\
\vec{\xi}_b&=&\frac{{\cal A}_bc_P}{A(0)}
  [g_S(\tm+\tp)+g_P\hat{k}\times(\tm-\tp)],
\end{eqnarray}
where ${\cal A}_{a,b}$ are defined in Eq.~(\ref{tanalpha}). In calculating
values for the polarizations, we have to give values to the anomalous coupling
constants $g_S$, $g_P$, $c_S$ and $c_P$. Scalar particle couplings arise in
many extensions of the SM. However, up to now there exist no definite
predictions about their values~\cite{Ananthanarayan:2003wi}. On the other
hand, the unparticle phenomenology stands beyond the other SM extension
models. Therefore, one has to make here quite voluntary presumptions that do
not lay on definite theoretical grounds. Here we use the ``SM-connected''
setting $g_S=g_V$, $g_P=g_A$, $c_S=c_V$, $c_P=c_A$, and $g_p=g_u=g$.

\begin{figure}
\begin{center}
\epsfig{file=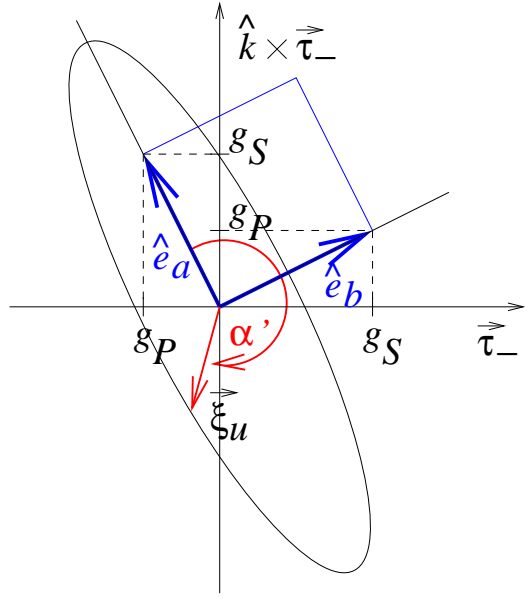, scale=0.8}
\caption{\label{figure4}The vector $\vec{\xi}_u$ in the plane spanned by
$\tm$ and $\hat k\times\tm$ for exemplary values for $c_S$ and $c_P$ and
arbitrary scale for the coefficients, where
$\tan\alpha'=({\cal A}_b/{\cal A}_a)\tan(d_u\pi)$.}
\end{center}
\end{figure}

\begin{figure}
\begin{center}
\epsfig{file=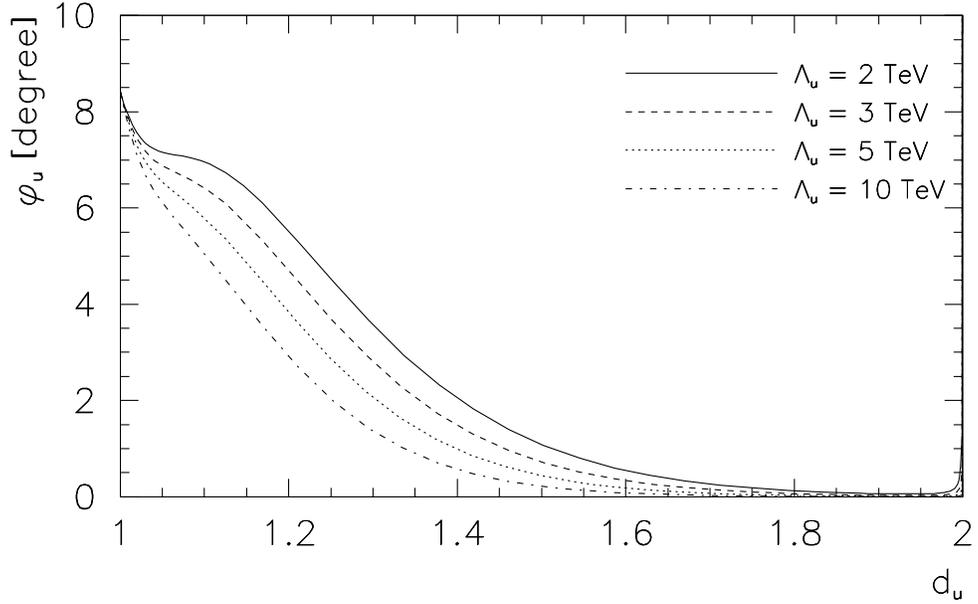, scale=0.8}
\caption{\label{figure5}Deviation angle $\varphi_u$ in dependence on $d_u$
for scales $\Lambda_u=2$, $3$, $5$, and $10\TeV$}
\end{center}
\end{figure}

\vspace{7pt}
The corrections to the top or antitop quark polarization due to anomalous
couplings are transverse to the top (antitop) quark polarizations due to the
SM which is antiparallel to the direction $\hat{k}$ of the initial beams. The
angle between these two components is
\begin{equation}\label{tanphi}
\tan\varphi_{p,u}=\frac{|\vec{\xi}_{p,u}|}{|\vec{\xi}_{SM}|}.
\end{equation}
Using $h_\pm=0$, $\tau_+=0$ and $\tau_-=0.8$ in order to eliminate the TP
dependent SM terms also close to the exact threshold, the vector $\vec{\xi}_u$
is a vector in the plane spanned by $\tm$ and $\hat k\times\tm$ orthogonal to
$\hat k$. However, a better reference frame to consider is the one spanned by
the orthonormal basis
\begin{equation}
\hat e_a=\frac{g_S\hat k\times\tm-g_P\tm}{\sqrt{g_S^2+g_P^2}|\tm|},\qquad
\hat e_b=\frac{g_S\tm+g_P\hat k\times\tm}{\sqrt{g_S^2+g_P^2}|\tm|}
\end{equation}
The situation is illustrated in Fig.~\ref{figure4}.
For different values of $d_u$ the vector runs on a ellipse with half axis
of length ${\cal A}_a\sqrt{g_S^2+g_P^2}|\tm|$ along $\hat e_a$ and half axis
of length ${\cal A}_b\sqrt{g_S^2+g_P^2}|\tm|$ along $\hat e_b$. The angle in
negative mathematical order with respect to $\hat e_a$ is given by $\alpha'$,
where $\tan\alpha'=({\cal A}_b/{\cal A}_a)\tan(d_u\pi)$. Together with the
$d_u$-dependence given by $K_u$ in Eq.~(\ref{tanphi}) we can calculate the
dependence of the deviation angle $\varphi_u$ on $d_u$ in the region
$1\le d_u<2$ for different values of the scale $\Lambda_u$. The result is
shown in Fig.~\ref{figure5}.
\begin{figure}
\begin{center}
\epsfig{file=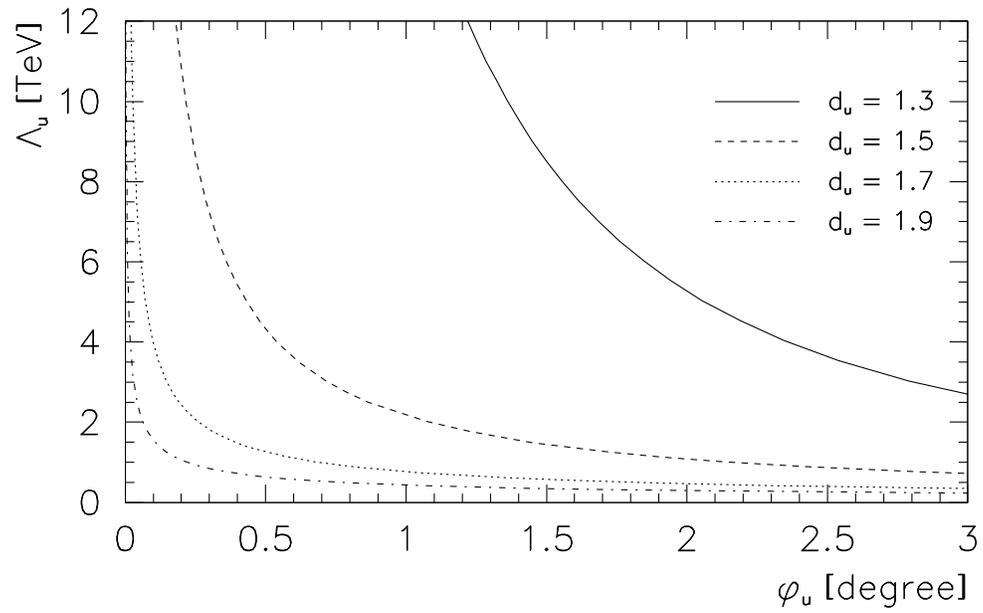, scale=0.8}
\caption{\label{figure6}Scale $\Lambda_u$ in dependence on the deviation
angle $\varphi_u$ for unparticle dimensions $d_u=1.3$, $1.5$, $1.7$, and $1.9$}
\end{center}
\end{figure}
\begin{figure}
\begin{center}
\epsfig{file=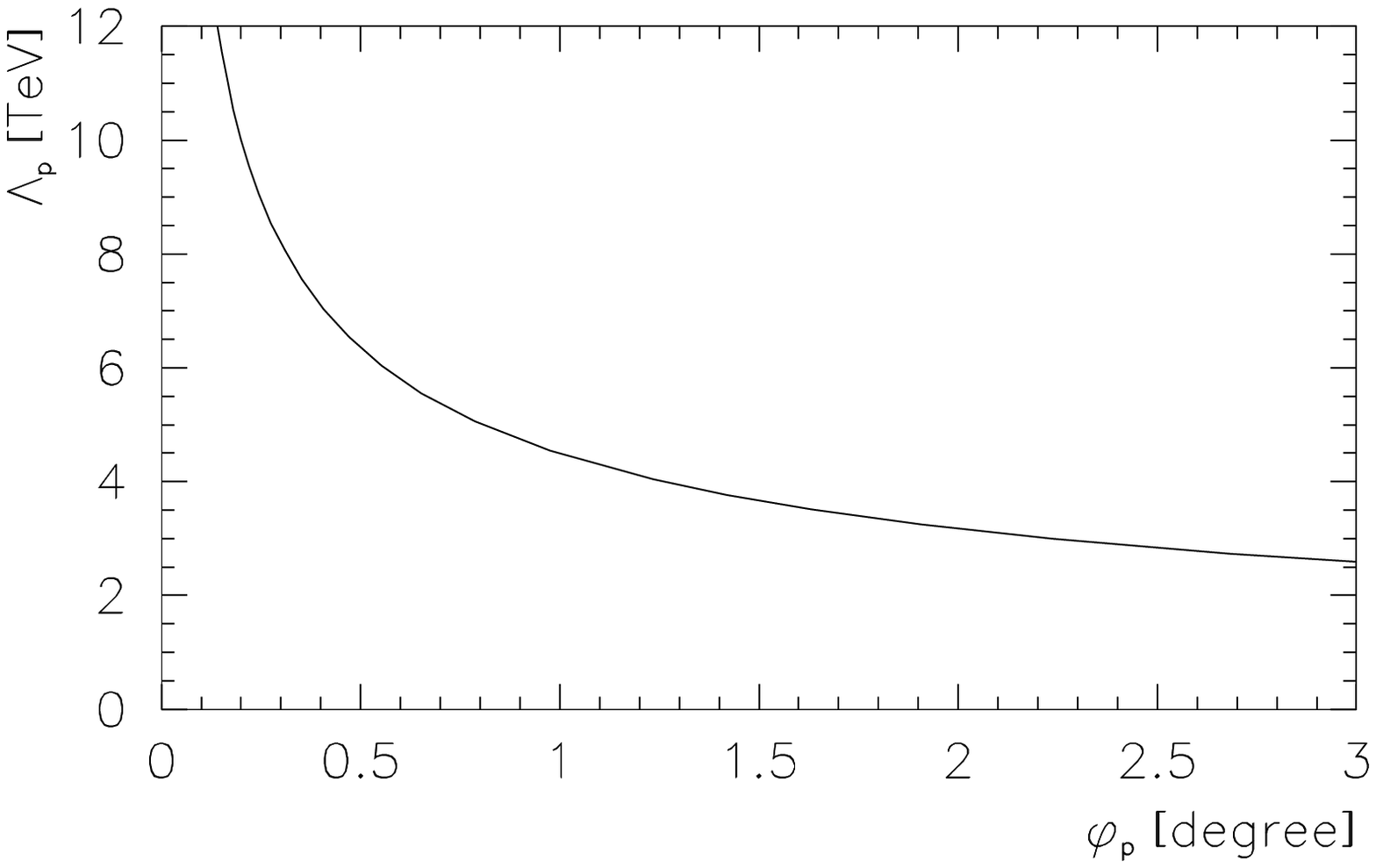, scale=0.8}
\caption{\label{figure7}Scale $\Lambda_p$ in dependence on the deviation
angle $\varphi_p$}
\end{center}
\end{figure}
The angle $\varphi_u$ turns out to be very sensitive to the dimension $d_u$,
increasing rapidly if $d_u$ is going to $d_u=1$. The fine structure of the
dependence close to $d_u=1.1$ is due to the elliptical dependence of the
vector $\vec{\xi}_u$ on $d_u$, superposed by the increase of $K_u$ for
$d_u\to 1$. Apparently, close to $d_u=1$ the value of the angle does no longer
depend on the scale $\Lambda_u$ but takes a constant value because, using
$\lim_{d_u\to 1}\sin(d_u\pi)\Gamma(1-d_u)=-\pi$, $K_{d_u=1}$ is given by
$K_1=-g_u^2/4k^2$. In Fig.~\ref{figure6} we show the dependence of the scale
$\Lambda_u$ on the deviation angle $\varphi_u$ for the values $d_u=1.3$, $1.5$,
$1.7$, and~$1.9$. Again, the strong dependence on $d_u$ is obvious. Assuming
that new physics is expected to appear at a scale of about $7\TeV$, the
detection of effects for $d_u=1.7$ and $1.9$ requires a high angle resolution
which may be not available for the near-future colliders.
On the other hand, anomalous scalar particle coupling effects for the same
assumed new-physics scale can be observed. This can be seen from
Fig.~\ref{figure7} where the dependence of the scale $\Lambda_p$ on the
deviation angle $\varphi_p$ is shown.

\section{Conclusion}
Our studies once more demonstrate the utility of using transversely polarized
initial beams for searching new-physics indications. The additional directions
provided by transverse polarization vectors can be successfully used for
constructing new measurable quantities both in the presence of final top
(antitop) polarization and its absence. In the previous case one loses
statistics but gains other advantages in separating anomalous coupling
signals from the SM contributions. The anomalous coupling contributions
depend linearly on the transverse polarization vectors. This circumstance
enables one to take only one of the initial beams to be transversely polarized.
Such a choice eliminates the transverse polarization depending SM
contributions. As an illustrative example we showed how to estimate the
anomalous scalar particle and unparticle coupling manifestations through the
measurement of the top quark polarization near the threshold of the process.

\subsection*{Acknowledgements}
The work is supported by the Estonian target financed projects No.~0180013s07
and No.~0180056s09 and by the Estonian Science Foundation under grant No.~6216.

\end{document}